\documentclass[conference,10pt]{IEEEtran_special}

\usepackage[cmex10]{amsmath}
\usepackage{cite}

\usepackage{latexsym}
\usepackage{color}
\usepackage{graphics}
\usepackage{amssymb}
\usepackage{amsthm}

\usepackage{cite}
\usepackage{array}
\usepackage{graphicx}

\usepackage{url}

\newcommand{\vect}[1]{\mathbf{#1}}

\def\Htran{\mbox{\tiny $\mathrm{H}$}}

\newcommand{\maximize}[1]{{\underset{{#1}}{\mathrm{maximize}}}}

\theoremstyle{plain}

\newtheorem{theorem}{Theorem}

\newtheorem{lemma}{Lemma}

\newtheorem{proposition}{Proposition}

\begin{document}

\title{Energy-Efficient Future Wireless Networks: \\ A Marriage between Massive MIMO and Small Cells\vspace{-1cm}}

\IEEEoverridecommandlockouts

\author{\IEEEauthorblockN{Emil Bj{\"o}rnson\IEEEauthorrefmark{1}, Luca Sanguinetti\IEEEauthorrefmark{2}\IEEEauthorrefmark{3}, and Marios Kountouris\IEEEauthorrefmark{4}}
\IEEEauthorblockA{\IEEEauthorrefmark{1}\small{Department of Electrical Engineering (ISY), Link\"{o}ping University, Link\"{o}ping, Sweden}}
\IEEEauthorblockA{\IEEEauthorrefmark{2}\small{Dipartimento di Ingegneria dell'Informazione, University of Pisa, Italy} $\quad$\IEEEauthorrefmark{3}\small{LANEAS, CentraleSup{\'e}lec, Gif-sur-Yvette, France}}
\IEEEauthorblockA{\IEEEauthorrefmark{4}\small{Mathematical and Algorithmic Sciences Lab, France Research Center, Huawei Technologies Co. Ltd.}
\thanks{This paper was supported in part by ELLIIT and the People Programme (Marie Curie Actions) FP7 PIEF-GA-2012-330731 Dense4Green.}%
}
}

\maketitle

\begin{abstract}
How would a cellular network designed for high energy efficiency look like?
To answer this fundamental question, we model cellular networks using stochastic geometry and optimize the energy efficiency with respect to the density of base stations, the number of antennas and users per cell, the transmit power levels, and the pilot reuse. The highest efficiency is neither achieved by a pure small-cell approach, nor by a pure massive MIMO solution. Interestingly, it is the combination of these approaches that provides the highest energy efficiency; small cells contributes by reducing the propagation losses while massive MIMO enables multiplexing of users with controlled interference.
\end{abstract}

\IEEEpeerreviewmaketitle

\vspace{-3mm}

\section{Introduction}
\label{sec:intro}

\vspace{1mm}

Two key goals for the fifth generation (5G) cellular networks are improved spectral efficiency (SE) and higher energy efficiency (EE) \cite{Andrews2014a}. These performance metrics are coupled and cannot be treated separately in the design of future networks \cite{Bjornson2014c}. The key to improve the SE and EE is higher spatial reuse; that is, more parallel transmissions per km$^2$. There are two main densification approaches: 1) smaller cell radius \cite{Hoydis2011c} and 2) massive MIMO (multiple input, multiple output) technology \cite{Larsson2014a}. The purpose of this paper is to show that these approaches are fundamentally non-competing; in fact, both are needed to make future wireless networks truly energy efficient.

The EE is defined as the benefit-cost ratio of the network:
\begin{equation*}
\mathrm{EE} \!=\! \frac{\textrm{Area Spectral Efficiency [bit/symbol/km}^2]}{\textrm{Transmit + Circuit Power per Area [Joule/symbol/km}^2]}
\end{equation*}
This definition reveals that there are three main factors that can be modified when optimizing the EE of a cellular network. 

Smaller cells are obtained by densifying the base station (BS) deployment. This has a positive effect on the EE in terms of increasing the SE and reducing the transmit power (due to lower propagation losses). However, the negative effect is the increased circuit power due the larger amount of hardware infrastructure. Small-cell deployments need to be modeled in an asymmetric fashion since the users are non-uniformly distributed over the coverage area. This makes Poisson point processes (PPPs) suitable and tractable analytic models \cite{Andrews2011a}.

In contrast, massive MIMO evolves the conventional BS technology by replacing the bulky high-gain vertical antennas with arrays comprising hundreds of small dipole antennas. Contrary to what the term ``massive'' suggests, massive MIMO arrays are rather compact; 160 dual-polarized antennas at 3.7 GHz fit into the form factor of a flat-screen television \cite{Vieira2014a}. By processing the antennas coherently, the array can receive and transmit a multitude of signals for users at different spatial locations. The positive effects on the EE comes from higher SE and less transmit power (due to an array gain), but the circuit power increases with the number of BS antennas.

Both approaches to improve EE are associated with nontrivial tradeoffs. In this paper, we optimize the deployment of a cellular network for higher EE on the uplink. The optimization variables are the BS density, the number of antennas and active users per BS, the transmit powers, and the pilot reuse. We prove the fundamental interplay between these parameters and show how to achieve high EE. Some previous works are the analytic single-cell analysis in \cite{Mohammed2014a,Bjornson2015a} and the numerical multi-cell analysis in \cite{Yang2013a,Bjornson2015a}. This is a continuation of our prior work in \cite{Bjornson2015c}, which was limited to perfect channel state information.

\begin{figure*}[t!]
 \begin{align} \label{eq:average-SINR} \tag{6}
  \underline {\mathrm{SINR}}  =
\frac{ M  }{ 
\Big( K + \frac{\sigma^2}{  \rho}  \Big) \! \Big(  1   + \frac{ 2}{\beta( \alpha-2)}  + \frac{\sigma^2}{\rho} \Big) \! + \!
 \frac{ 2K}{\alpha-2} \! \left(  1  + \frac{\sigma^2}{\rho} \right) \! + \!
\frac{K}{\beta} \! \left( \frac{4}{(\alpha-2)^2 } +
\frac{1 }{ \alpha-1} \right) \!
+  M \frac{ 1}{\beta( \alpha-1)}  }
\end{align}
\hrulefill \vspace{-4mm}
\end{figure*}

\section{System Model}

\vspace{1mm}

We consider a cellular network that is designed to serve users from a heterogeneous user distribution. 
To this end, we adopt the stochastic geometry model in \cite{Andrews2011a}, where the BSs are distributed in $\mathbb{R}^2$ according to a homogeneous PPP $\Phi_{\lambda}$ of intensity $\lambda$ [BSs per $\textrm{km}^2$]. More precisely, this means that in any area of size $A$ $\textrm{km}^2$, the number of BSs is a random number from a Poisson distribution with mean value $\lambda A$. These BSs are uniformly and independently distributed over the area.

\begin{figure}
\begin{center}
\includegraphics[width=\columnwidth]{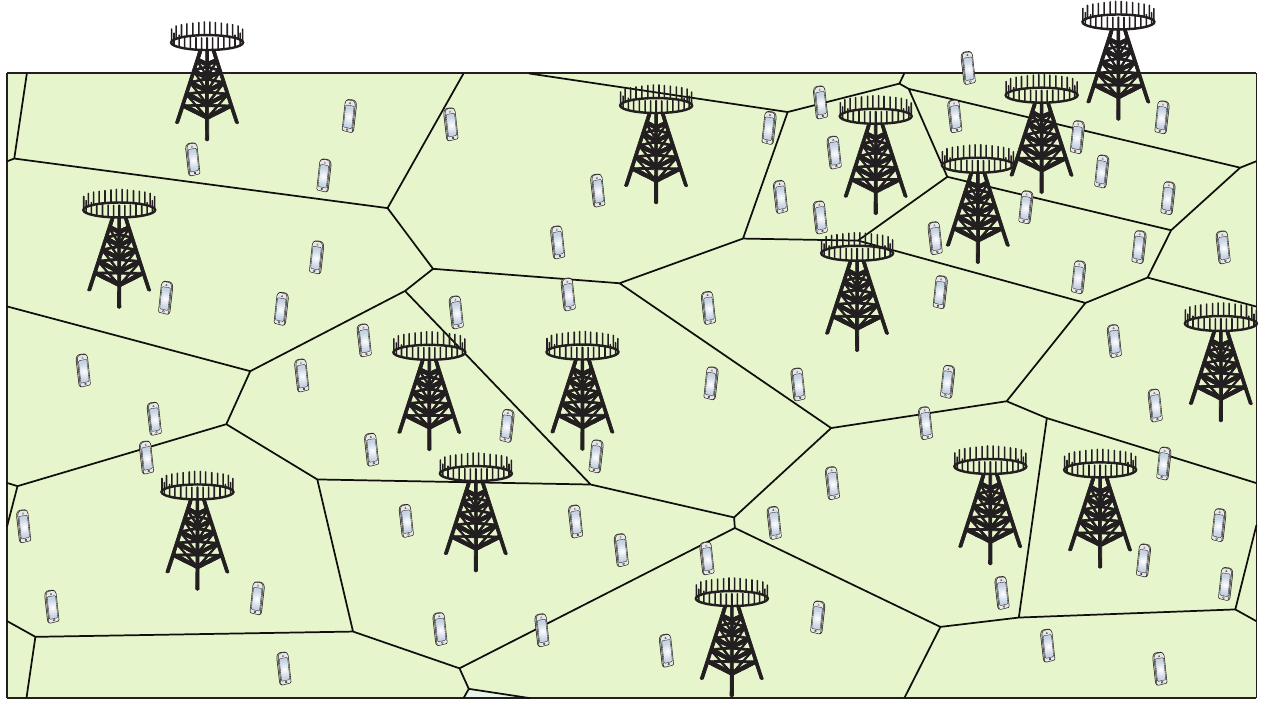}
\end{center}\vskip-5mm
\caption{Illustration of one realization of BS positions from the PPP $\Psi_{\lambda}$ with $K$ users uniformly distributed in the corresponding Poisson-Voronoi cells.} \label{figureVoronoi} \vskip-4mm
\end{figure}

Each BS has $M$ antennas and serves $K$ single-antenna user equipments (UEs). Each UE connects to its closest BS, hence the coverage area of a BS is its Poisson-Voronoi cell; see Fig.~\ref{figureVoronoi}. The UEs are assumed to be uniformly distributed in the cells. Since there are $K$ UEs per cell, small cells have higher user density (per $\textrm{km}^2$) than larger cells, giving a judicious deployment based on a heterogenous user distribution.\footnote{Alternatively, the UEs could have been distributed according to an independent homogeneous PPP. This is less sensible since smaller cells would on average have fewer UEs than larger cells, which contradicts the main principle of densifying networks mainly at places with high user loads.}

We consider uplink transmission where the time-frequency resources are divided into blocks of $T_c$ seconds and $W_c$ Hz. This leaves room for a total number of $S = T_c W_c$ transmission symbols for pilots and data. The channel response between $\mathrm{BS}_l  \in \Phi_{\lambda}$ and UE $k$ in cell $j$ is modeled in each block as a random Rayleigh fading vector $\vect{h}_{ljk} \sim \mathcal{CN}(\vect{0}, \omega^{-1} d_{ljk}^{-\alpha} \vect{I}_M)$. The variance is given by the Euclidean distance $d_{ljk}$ [km], the pathloss exponent $\alpha> 2$, and the parameter $\omega$ that determines the propagation loss at the reference distance of 1 $\mathrm{km}$.

The translation invariance of PPPs allows to perform statistical performance analysis for a \emph{typical UE} at the origin, which is statistically representative for any other UE in the network \cite{Baccelli2008a}. Assume that this typical UE has the arbitrary index $k$ and is connected to a typical BS, denoted as $\mathrm{BS}_0 \in \Phi_{\lambda}$. 
Then, the following basic properties hold (see for example \cite{Weber2010a}):

\begin{lemma} \label{lemma:distance-distribution}
The distance from the typical UE to its serving BS is $d_{00k} \sim \mathrm{Rayleigh} \big( \frac{1}{\sqrt{2\pi \lambda}} \big)$. The BSs of the other cells form a homogeneous PPP $\Psi_{\lambda} = \Phi_{\lambda} \!\setminus \! \{ \mathrm{BS}_0 \}$ in $\{ \vect{x} \in \mathbb{R}^2 : \| \vect{x} \| \!\geq\! d_{00k} \}$.
\end{lemma}

The received signal $\vect{y}_0 \in \mathbb{C}^{M}$ at $\mathrm{BS}_0$ is modeled as
\begin{equation} \label{eq:system-model-received}
\vect{y}_0 = \sum_{i=1}^{K} \sqrt{p_{0i}} s_{0i} \vect{h}_{00i}  + \sum_{j \in \Psi_{\lambda}}
\sum_{i=1}^{K} \sqrt{p_{ji}} s_{ji}  \vect{h}_{0ji}  + \vect{n}_{0}\vspace{-2mm}
\end{equation}
where $\vect{n}_{0} \sim \mathcal{CN}(0,\sigma^2 \vect{I}_M)$ accounts for the receiver noise. 
The arbitrary symbol transmitted by UE $i$ in cell $j$ is denoted by $s_{ji} \in \mathbb{C}$ and is normalized to unit power (i.e., $\mathbb{E}\{ | s_{ji}|^2 \}=1$), while the corresponding transmit power $p_{ji}  \geq 0$ is selected as
$p_{ji} = \rho  \omega d_{jji}^{\alpha}$
with $\rho \geq 0$ being a power-control coefficient that will be optimized later on. 
This is a power-control policy based on statistical channel-inversion that gives an average effective channel gain of $\mathbb{E}\left\{{p_{ji} \|\vect{h}_{jji}\|^2} \right\} = M {\rho}$ for all UEs (irrespective of their locations), which is necessary to avoid near-far blockage in uplink multi-user MIMO systems.

Pilot transmission is used to estimate channels. We assume that there are $B$ pilot symbols per block, where $K \leq B \leq S$. Each BS selects $K$ different pilots uniformly at random in each block and we call $\beta = \frac{B}{K} \ge 1$ the \emph{pilot reuse factor}. Hence, the probability that $\mathrm{BS}_j$ has a UE (with index $k$ for convenience) that reuses the pilot of the typical UE is $1/\beta$. Let $\chi_{0kj} \in \{0,1\}$ be the corresponding random variable, where $\chi_{0kj} =1$ has probability $1/\beta$ and means that cell $j$ reuses the pilot of the typical UE, then $\mathrm{BS}_0$ receives the signal
\begin{equation} \label{eq:received-pilot}
\vect{z}_{0k} = \sqrt{p_{0k}} \vect{h}_{00k} + \sum_{j \in \Psi_{\lambda} } \chi_{0kj} \sqrt{p_{jk}} \vect{h}_{0jk} + {\bf{n}}_{0}
\end{equation}
during pilot transmission from the typical UE. The minimum mean-squared error (MMSE) estimate of $\vect{h}_{00k}$ is \cite{Bjornson2016b} \vspace{-1mm}
\begin{equation} \label{eq:MMSE-estimator}
  \hat{\vect{h}}_{00k} =  \sqrt{ \frac{1}{\rho \omega d_{00k}^{\alpha}} } 
\frac{ 1}{ 1   +  \sum_{j \in \Psi_{\lambda} }    \chi_{0kj}    \frac{d_{jjk}^{\alpha}}{ d_{0jk}^{\alpha} } + \frac{\sigma^2}{ \rho} } \vect{z}_{0k}\vspace{-2mm}
\end{equation}
and the estimation error $\Delta\vect{h}_{00k} = \vect{h}_{00k}  - \hat{\vect{h}}_{00k}$ is distributed as  $\Delta\vect{h}_{00k}  \sim\mathcal{CN}(\vect{0},\vect{C}_{00k})$ with 
\begin{equation} \label{eq:MMSE-error-cov}
\vect{C}_{00k} = \frac{1}{\omega d_{00k}^{\alpha}}  \Bigg(1 - \frac{  1 }{ 1   +  \sum_{j \in \Psi_{\lambda} }    \chi_{0kj}   \frac{d_{jjk}^{\alpha}}{ d_{0jk}^{\alpha} } + \frac{\sigma^2}{\rho} }  \Bigg) \vect{I}_M.
\end{equation}

We assume that the BSs use the estimated channels for maximum ratio combining (MRC). In particular, the data symbols transmitted by the typical UE are detected at $\mathrm{BS}_0$ by correlating the received signal in
\eqref{eq:system-model-received} with the MMSE estimate as $r_{0k} = \nu_{00k}\hat{\vect{h}}_{00k}^{\Htran}  \vect{y}_0$, where $\nu_{00k} \in \mathbb{C}$ is a scaling factor \cite{Bjornson2016b}.

\section{Problem Formulation}
\label{sec:ASE}

\vspace{1mm}

In this section, we formulate the EE maximization problem considered in this paper. Since the ergodic capacity of a network with imperfect channel knowledge and inter-cell interference modeled as a shot-noise process is not known yet \cite{Weber2010a}, the first step towards formulating the EE is thus to obtain a tractable SE expression. We derive the following lower bound using the approach from \cite{Jose2011b} and Jensen's inequality.

\begin{proposition} \label{prop:average-SE}
If MRC is employed, a lower bound on the UL average SE [bit/symbol/user] is
\begin{equation} 
\underline {\mathrm{SE}} = \Big( 1 - \frac{\beta K}{S} \Big)\! \log_2 \! \left( 1 + \underline {\mathrm{SINR}} \right)\label{eq:average-SE}
\end{equation}
where $\underline {\mathrm{SINR}} $ is given by \eqref{eq:average-SINR} at the top of the page.\end{proposition}

\setcounter{equation}{6}

This lower SE bound holds for any $\beta\geq 1$ such that $\beta K \leq S$, since the pilot signals need to be contained in a coherence block. Observe that $\beta K$ does not need to be an integer since an arbitrary $\beta K$ can be achieved by switching (for appropriate fractions of time) between the closest smaller integer and the closest larger integer.

This work aims at maximizing the EE with respect to the optimization variables $\theta = (\beta, \rho,\lambda,K,M)$. The area spectral efficiency (ASE) [bit/symbol/$\textrm{km}^2$] is computed as
\begin{equation}
\mathrm{ASE}(\theta) = \lambda K \Big( 1 - \frac{\beta K}{S} \Big)\! \log_2 \! \left( 1 + \underline {\mathrm{SINR}} \right).\label{eq:def-ASE}
\end{equation}
The area energy consumption (AEC) [Joule/symbol/$\textrm{km}^2$] accounts for radiated signal energy, dissipation in analog hardware, digital signal processing, backhaul signaling, and various overhead costs (e.g., cooling and power supply losses). A detailed model of these factors was given in \cite{Bjornson2015a}:
\begin{align} \label{eq:def-AEC}
&\mathrm{AEC}(\theta) = \\ &\! \lambda \left(  \frac{S \!-\! \beta K \!+\! 1}{S}  \frac{\rho \omega}{\eta}  \frac{ \Gamma(\frac{\alpha}{2} +1 ) }{ (\pi \lambda)^{\alpha/2} } K \!+\! \mathcal{C}_0 \!+\! \mathcal{C}_1 K \!+\! \mathcal{D}_0 M \!+\! \mathcal{D}_1 M K \right)  \notag
\end{align}
where we have used that 
\begin{align}\label{avg_p_ji}
\mathbb{E} \{ p_{ji} \} =  \rho  \omega \frac{ \Gamma({\alpha}/{2} +1 ) }{(\pi \lambda)^{\alpha/2}} 
\end{align}
and that each UE transmits $S - \beta K$ data symbols and one pilot symbol per block \cite{Bjornson2016b}.
The amplifier efficiency is $\eta \in (0,1]$, $\mathcal{C}_0$ models the static energy consumption of a BS, and $\mathcal{D}_0 M$ models the energy consumption of the BS transceiver chains.
Moreover, $\mathcal{C}_1 K + \mathcal{D}_1 M K$ accounts for the energy consumed by a UE and the BS signal processing. The forthcoming analysis holds for any positive value of these parameters, but examples are given in Table \ref{table_parameters_hardware}.

The objective of this paper is to solve the following constrained EE maximization problem:
\begin{equation} \label{eq:main-optimization-problem}
\begin{split}
\maximize{\theta \in \Theta}\, &\quad \mathrm{EE}(\theta) = \frac{\mathrm{ASE}(\theta)}{\mathrm{AEC}(\theta)}\\
\mathrm{subject} \,\, \mathrm{to}  & \quad \underline{\mathrm{SINR}} \ge \gamma 
\end{split}
\end{equation}
where $\Theta = \{\theta: \, \rho\geq 0, \lambda \geq 0, \beta \geq 1, (M,K) \!\in\! \mathbb{Z}_+, K\beta \leq S\}
$ is the set of feasible optimization variables and the parameter $\gamma>0$ imposes an average SE constraint of $\log_2(1+\gamma)$ [bit/symbol/user], where the average is computed with respect to both BS and UE locations.
The need for such a constraint comes from the observation that an unconstrained EE maximization may lead to very low SEs per UE.

\section{Energy-Efficiency Maximization}
\label{sec:maximization}

\vspace{1mm}
In this section, we solve the EE maximization problem in \eqref{eq:main-optimization-problem}. The proofs of all theorems are omitted due to space limitations but can be found in \cite{Bjornson2016b}.

\subsection{Optimal Pilot Reuse Factor $\beta$}

\vspace{1mm}

We begin by deriving the optimal value of the pilot reuse factor $\beta$ when the other optimization variables are fixed. As stated in the following theorem, the EE-optimal value of $\beta$ is such that the SINR constraint is met with equality.

\begin{theorem} \label{th:optimal-beta}
Consider any set $\{\rho, \lambda, M, K\}$ for which the problem \eqref{eq:main-optimization-problem} is feasible for some $\beta$. The SINR is an increasing function of $\beta$ and the SINR constraint in \eqref{eq:main-optimization-problem} is satisfied by selecting 
\begin{equation} \label{eq:beta-optimal}
\beta^\star = \frac{B_1 \gamma}{M - B_2 \gamma}
\end{equation}
where
\begin{align} \label{eq:B1}
B_1 &= \frac{4K}{(\alpha-2)^2 } +
\frac{K +M }{ \alpha-1} +  \frac{ 2 ( K + \frac{\sigma^2}{  \rho} )}{ \alpha-2}  \\
B_2 &=  \left( K + \frac{\sigma^2}{  \rho} +  \frac{ 2K}{\alpha-2} \right) \left(  1  + \frac{\sigma^2}{\rho} \right). \label{eq:B2}
\end{align}
\end{theorem}

This theorem explains how the EE-optimal pilot reuse factor $\beta^\star$ depends on the other system parameters. Recall that larger $\beta$ leads to higher estimation accuracy. From \eqref{eq:beta-optimal}--\eqref{eq:B2}, it is easily seen that, to guarantee a certain SINR, $\beta^\star$ increases with $K$ (since the interference increases with more UEs) whereas $\beta^\star$ decreases with $\rho$ as well as with $M$ (as it follows from taking the derivative of $\beta^\star$ with respect to $M$). This is because both $\rho$ and $M$ amplify the desired signal, which as consequence improves channel estimation and makes the system operate in a less noise limited regime.

\subsection{Optimal BS Density and Radiated Power}

\vspace{1mm}

Based on $\beta^\star$ in Theorem \ref{th:optimal-beta}, the optimal values for the BS density $\lambda$ and the power control coefficient $\rho$ are as follows.

\begin{theorem} \label{th:optimal-lambda}
Define $\rho = \lambda \tilde{\rho}$ for $\tilde{\rho}>0$ and consider any set of $\{\tilde{\rho},M,K\}$ for which the problem \eqref{eq:main-optimization-problem} is feasible using $\beta^{\star}$. The EE is then a monotonically increasing function of $\lambda$ and is maximized as $\lambda \to \infty$.
\end{theorem}

This theorem proves that from an EE perspective it is preferable to have as high BS density as possible. This might sound counterintuitive at first, since the inter-cell interference typically grows as the cells become smaller. However, for any $\alpha >2$ the adopted power control policy reduces the average transmit power as $\tilde{\rho}/\lambda^{\alpha/2-1}$ with the BS density $\lambda$ (as it follows from \eqref{avg_p_ji}). Thus, the desired and interfering signals are amplified equally much on average; in other words, it is only the impact of noise that vanishes as $\lambda$ increases.

It is clearly unreasonable to have an infinitely high BS density, but the numerical results in Section \ref{sec:numerical} show that the asymptotic limit is approached already at $\lambda = 10$ BS/km$^2$, which is a modest number. In ultra-dense deployments, $\lambda$ is expected to be one or several orders of magnitude larger than this. Hence, we can proceed the EE maximization analysis by actually letting $\lambda \to \infty$ without loss of realism. 

\subsection{Optimal Number of Antennas and UEs per BS}
\label{subsec:opt-M-K}

\vspace{1mm}

By using Theorems \ref{th:optimal-beta} and \ref{th:optimal-lambda}, the original EE maximization problem in  \eqref{eq:main-optimization-problem} has been reduced to\footnote{
We have also dropped the constraint $\beta^\star \leq S/K$ since the EE becomes negative when this is not satisfied, thus the maximal EE is not affected.}
\begin{align} \label{eq:main-optimization-problem-modified2}
\maximize{M,K \in \mathbb{Z}_+} &\quad \frac{ K (1 - \frac{K}{S} \frac{\bar{B}_1 \gamma}{M  -\bar{B}_2 \gamma} )  \log_2(1+\gamma)}{  \mathcal{C}_0 + \mathcal{C}_1 K + \mathcal{D}_0 M + \mathcal{D}_1 M K   } \\
\,\,\,\, \,\, \mathrm{subject} \,\, \mathrm{to} \,\,\, & \quad  \frac{\bar{B}_1 \gamma}{M -\bar{B}_2 \gamma}  \geq 1 \label{eq:main-optimization-problem-modified2-constraint}
\end{align}
where we have defined (as obtained from \eqref{eq:B1} and \eqref{eq:B2} when $\rho \to \infty$)
\begin{align} \label{eq:B1bar}
\bar{B}_1 &= K \left( \frac{4}{(\alpha-2)^2 } +
\frac{1}{ \alpha-1} +  \frac{ 2}{ \alpha-2} \right)  +
\frac{M  }{ \alpha-1}  \\
\bar{B}_2 &=  K\left(  1 +  \frac{ 2}{\alpha-2} \right) . \label{eq:B2bar}
\end{align}
To find the optimal values for $M$ and $K$, an integer-relaxed version of \eqref{eq:main-optimization-problem-modified2} is first considered where $M$ and $K$ can be any positive real-valued scalars. The integer-valued solutions are extracted from the relaxed problem. For analytic tractability, we first fix the number of BS antennas per UE, $\bar{c} = {M}/{K}$, and find the EE-maximizing value of $K$ as follows.

\begin{theorem} \label{th:optimal-K}
Consider the optimization problem \eqref{eq:main-optimization-problem-modified2} where $M$ and $K$ are relaxed to be real-valued variables. For any fixed $\bar{c} = {M}/{K}> 0$ such that the relaxed problem is feasible for some $K$, the EE is maximized by
\begin{align} \label{eq:optimal-K}
K^{\star} = \frac{ \sqrt{\left(G\mathcal C_{0}\right)^{2}+\mathcal{C}_{0} \mathcal D_{1}\bar c + \mathcal{C}_{0} G \left(\mathcal C_{1} + \mathcal D_{0}\bar c \right)}   - G\mathcal C_{0} }{\mathcal D_{1}\bar c+ G\left(\mathcal C_{1} + \mathcal D_{0}\bar c \right)}
\end{align} 
where
\begin{align} 
G = \frac{1}{S} \frac{ \frac{4\gamma}{(\alpha-2)^2 } +
\frac{\gamma}{ \alpha-1} +  \frac{ 2\gamma}{ \alpha-2}   + \frac{ \gamma }{ \alpha-1}  \bar{c} }{ 
  \bar{c}
  - \left(  1 +  \frac{ 2}{\alpha-2} \right) \gamma }.
\end{align}
\end{theorem}

Similarly, if $K$ is fixed, then the EE-maximizing value of  $M$ is obtained as follows.

\begin{figure*}[t!]
\begin{align} \label{eq:c1}
M^\star=  K \frac{a_1 K  + a_2  + \sqrt{  a_1  a_2 K  + a_1^2 K^2 + 
( 1 - a_0 K   )( a_1 K +a_0  a_2 K )
\frac{ \mathcal{C}_0 + \mathcal{C}_1 K}{\mathcal{D}_0 K + \mathcal{D}_1 K^2} 
 + a_0 a_1 a_2 K^2 + a_0 a_2^2 K } }{ 1 - a_0 K }
\end{align}
\hrulefill \vspace{-4mm}
\end{figure*}

\begin{theorem} \label{th:optimal-M}
Consider the optimization problem \eqref{eq:main-optimization-problem-modified2} where $M$ and $K$ are relaxed to be real-valued variables. For any fixed $K> 0$ such that the relaxed problem is feasible, the EE is maximized by
$M^\star$ in \eqref{eq:c1}, at the top of the next page, if it satisfies the constraint in \eqref{eq:main-optimization-problem-modified2-constraint}. Otherwise, it is maximized by
\begin{align} \label{eq:c2}
M^\star= \frac{ K \gamma \left( 1 +\frac{4}{(\alpha-2)^2 } + \frac{1}{ \alpha-1} +  \frac{ 4}{ \alpha-2}   \right)}{  1 - \frac{ \gamma }{\alpha-1}  }.
\end{align}
The following parameters are used in these expressions:
\begin{align}
a_0 &= \frac{ \gamma }{S(\alpha-1)} \\ 
a_1 &=  \frac{1}{S} \left( \frac{4\gamma}{(\alpha-2)^2 } + \frac{\gamma}{ \alpha-1} +  \frac{ 2\gamma}{ \alpha-2}   \right) \\
a_2 & = \left(  1 +  \frac{ 2}{\alpha-2} \right) \gamma .
\end{align}
\end{theorem}

Theorems  \ref{th:optimal-K} and \ref{th:optimal-M} show how $K$ and $M$ are related at the EE-optimal points.
We notice that that $M^{\star}$ increases with $K$, and vice versa.  An important question is whether the EE is maximized at $K=1$ (single-user transmission) or at $K >1$ (multi-user transmission). The answer is ultimately determined by the hardware characteristics: From \eqref{eq:optimal-K} it is found that $K^{\star}$ increases with the static energy consumption $\mathcal{C}_0$, while it decreases with $\mathcal{C}_1$, $\mathcal{D}_0$, and $\mathcal{D}_1$ that are the terms of the AEC that increase with $K$ and $M$. Similarly, $M^{\star}$ increases with $\mathcal{C}_0$ and $\mathcal{C}_1$, but decreases with $\mathcal{D}_0$ and $\mathcal{D}_1$. The intuition is that more hardware should be turned on (i.e., BS antennas and UEs per cell) only if the increase in circuit power has a marginal effect on the total AEC. Similarly, with a larger static consumption we can afford to turn on more BS antennas and UEs since the relative power cost is lower.
In addition, we note that $M^*$ is an increasing function of $\gamma$, since deploying more antennas is a reliable way to achieve higher rates.

We can now devise an alternating optimization algorithm that solves the integer-relaxed EE maximization problem:

\begin{enumerate}
\item Find a feasible starting point $(M,K)$ to \eqref{eq:main-optimization-problem-modified2};
\item Optimize $K$ for a fixed $\bar{c}=M/K$ using Theorem \ref{th:optimal-K};
\item Optimize $M$ for a fixed $K$ using Theorem \ref{th:optimal-M};
\item Repeat 2)--3) until convergence is achieved.
\end{enumerate}

This algorithm can be shown to converge to the global optimal solution $(K^{\star \star},M^{\star \star}) \in \mathbb{R}^2$ of the integer-relaxed version of the EE maximization problem \eqref{eq:main-optimization-problem-modified2}; see \cite{Bjornson2016b} for details. It is hard to find an equally strong result for the integer-valued global optimum to \eqref{eq:main-optimization-problem-modified2}, but $(K^{\star \star},M^{\star \star})$ is a good starting point for finding the best integers $K$ and $M$. In fact, one can show that the EE is quasi-concave with respect to $K$ and $M$, which implies that the integer-solution is contained in a convex level set around $(K^{\star \star},M^{\star \star})$, which effectively limits the distance from the solution to the relaxed problem.

To summarize, the EE maximization problem in \eqref{eq:main-optimization-problem} was solved by selecting $\beta$ such the SINR constraint is satisfied with equality, letting $\lambda \to \infty$ (which removes the impact of $\rho$), and devising an alternating optimization algorithm that provides the real-valued $K$ and $M$ that maximize the EE. The final solution  $(K^{\star},M^{\star})$ is then found by  searching through the integer points in the vicinity of the real-valued solution, and capitalizing on the quasi-concavity. In the process of solving \eqref{eq:main-optimization-problem}, Theorems \ref{th:optimal-beta}--\ref{th:optimal-M} have also exposed the fundamental interplay between the optimization variables and how they depend on the hardware characteristics and propagation parameters.

\section{Numerical Examples}
\label{sec:numerical}

\vspace{1mm}

In this section, we provide numerical results that illustrate and validate the analytic results of previous sections. The optimization problem in \eqref{eq:main-optimization-problem} depends on a number of parameters that describe the propagation and hardware characteristics. Table \ref{table_parameters_hardware} provides the parameter values used in the simulations.

\begin{table}[!t]
\renewcommand{\arraystretch}{1.3}
\caption{Simulation Parameters}
\label{table_parameters_hardware}  \vspace{-1mm}
\centering
\begin{tabular}{|c|c|c|}
\hline
\bfseries Parameter & \bfseries Symbol & \bfseries Value \\
\hline

Pathloss exponent & $\alpha$ & $3.76$ \\

Coherence block length & $S$ & $400$ \\

Propagation loss at 1 $\textrm{km}$ & $\omega$ & $130 \, \mathrm{dB}$ \\

Power amplifier efficiency & $\eta$ & $0.39$ \\

Symbol time & $\tau$ & $\frac{1}{2 \cdot 10^7} \,\, \mathrm{[s/symbol]}$ \\

Static energy consumption & $\mathcal{C}_0$ & $10 \, \mathrm{W} \cdot \tau \,\, \mathrm{[J/symbol]}$ \\

Circuit energy per active user &  $\mathcal{C}_1$ & $0.1 \, \mathrm{W} \cdot \tau \,\, \mathrm{[J/symbol]}$ \\

Circuit energy per BS antenna & $\mathcal{D}_0$ & $1 \, \mathrm{W} \cdot \tau \,\, \mathrm{[J/symbol]}$ \\

Signal processing coefficient & $\mathcal{D}_1$ & $1.56 \cdot 10^{-10} \, \mathrm{[J/symbol]}$ \\

Noise variance & $\sigma^2$ & $10^{-20} \, \mathrm{[J/symbol]}$ \\

\hline
\end{tabular} \vskip-1mm
\end{table}

\begin{figure} \vspace{-1mm}
\begin{center}
\includegraphics[width=\columnwidth]{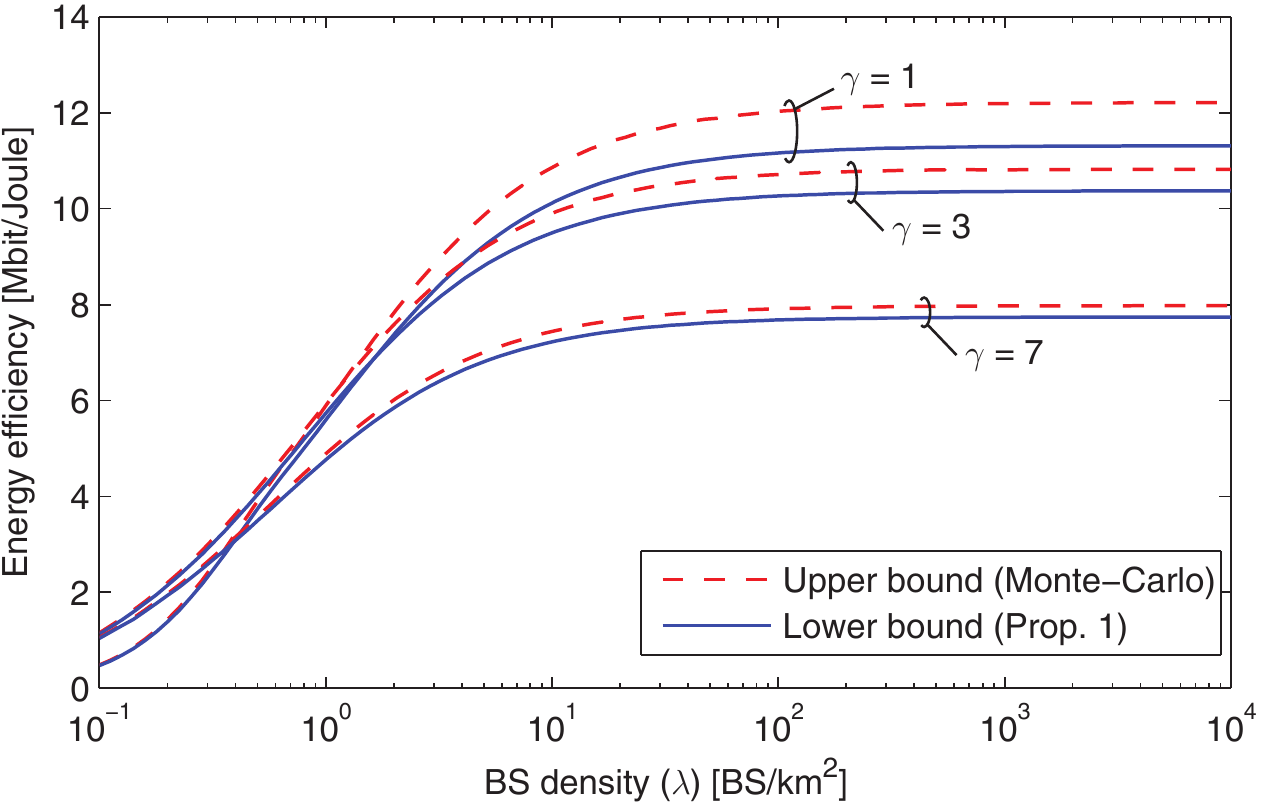}
\end{center}\vskip-4mm
\caption{Optimized Energy efficiency (in Mbit/Joule) as a function of the BS density for different SINR constraints.} \label{figureBSdensity} \vskip-4mm
\end{figure}

Fig.~\ref{figureBSdensity} shows the EE as a function of $\lambda$, when the other variables are optimized numerically for each given value of $\lambda$. We show results for the lower bound on the SE in Proposition \ref{prop:average-SE} and an upper bound obtained by numerical averaging over the BS and UE positions for a finite number of interferers.
We consider the SINR constraints $\gamma \in \{  1, \, 3, \, 7 \}$ and in all three cases there is only a small gap between the lower and upper bounds. The important message from Fig.~\ref{figureBSdensity} is that the EE increases with $\lambda$, but begins to saturate around $\lambda = 10$. Hence, EE maximization based on letting $\lambda \rightarrow \infty$ gives realistic results in both contemporary and future urban deployments.

\begin{figure}[t!]
\begin{center}
\includegraphics[width=\columnwidth]{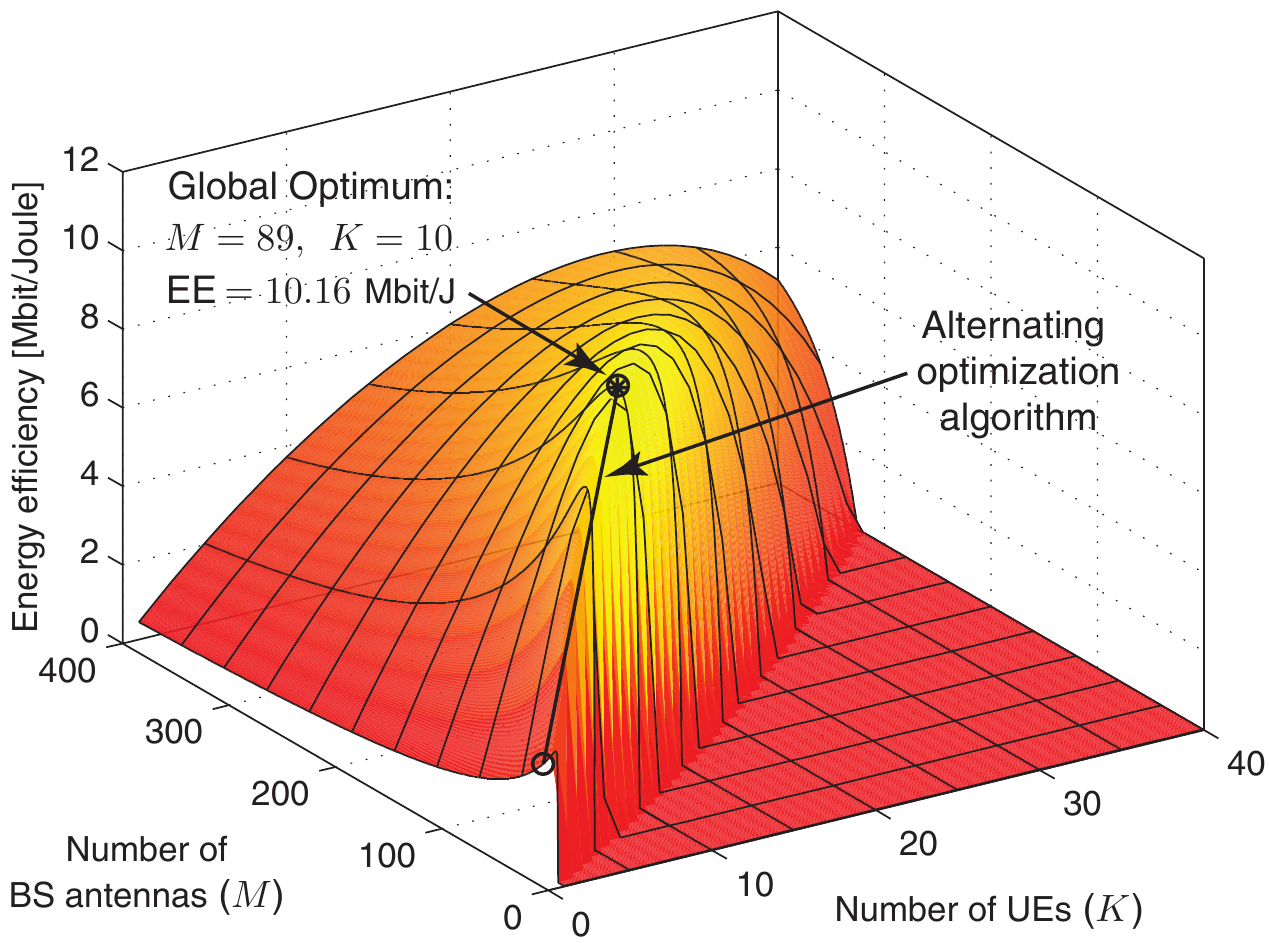}
\end{center}\vskip-4mm
\caption{Energy efficiency (in Mbit/Joule) for $\gamma = 3$. The optimum is star-marked and the convergence
of the alternating algorithm is shown with circles.} \label{figure3d}  \vskip-3mm
\end{figure}

Fig.~\ref{figure3d} shows the EE for $\gamma = 3$ as a function of $M$ and $K$, when $\lambda$, $\rho$, and $\beta$ are optimized. The maximal EE is $10.156$ Mbit/Joule and is achieved by $(M^\star,K^\star,\beta^*) = (89,10,7.24)$. This solution is a massive MIMO configuration with a high pilot reuse factor.  
The rationale is that large $M$ and $\beta$ give good interference suppression, while the large $K$ is a way to share the energy costs associated with a BS between UEs.

The alternating optimization algorithm is also illustrated in Fig.~\ref{figure3d}. It is initiated at $(M,K) = (20,1)$ and converges in three iterations to the real-valued solution $(M^{\star \star},K^{\star \star}) = (91.2,10.2)$ with an EE of $10.375$ Mbit/Joule, which is $0.013 \%$ higher than the EE at the integer-valued solution. Hence, the EE performance is quite flat around the global optimum; it decreases quickly with $K$ but slowly with $M$.

\subsection{EE Maximization for a Given UE Density}

\vspace{1mm}

Next, we study the tradeoff between massive MIMO and small cells when a cellular network is deployed to cover a heterogeneous UE density of $\mu$ UE/km$^2$. Mathematically, this amounts to solving \eqref{eq:main-optimization-problem} with the additional constraint $\mu = K \lambda$,
which can be easily solved numerically. We consider the range $\mu \in [10^2, 10^5]$ since the METIS project report\cite{METIS_D11_short} predicts future UE densities from $\mu=10^2$ UE/km$^2$ (in rural areas) to $\mu=10^5$ UE/km$^2$ (in shopping malls).

\begin{figure}[t!]
\begin{center}
\includegraphics[width=\columnwidth]{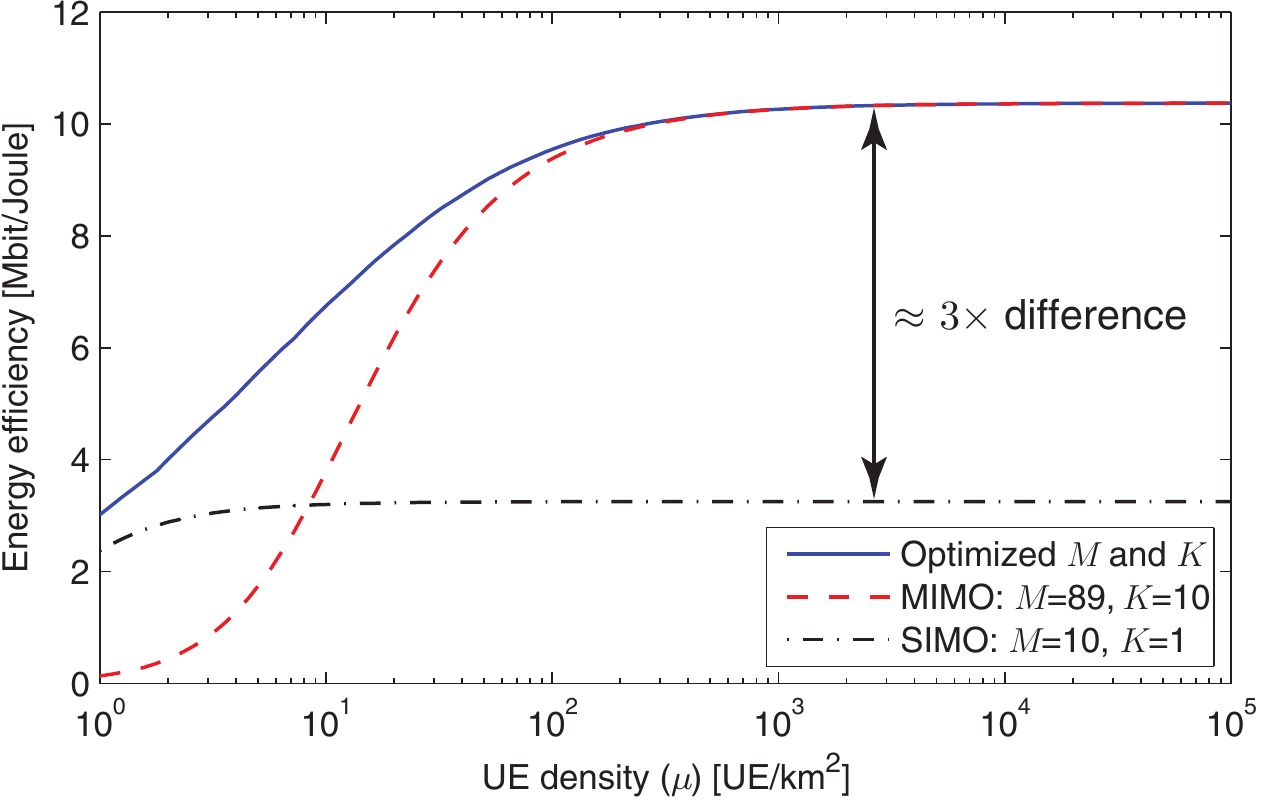}
\end{center}\vskip-4mm
\caption{Energy efficiency (in Mbit/Joule) as a function of the UE density $\mu$. The EE is optimized according to \eqref{eq:main-optimization-problem} with the extra constraint $\mu = K \lambda$, or only with respect to $(\lambda,\beta,\rho)$ for given $M$ and $K$ in the reference cases.} \label{figureUEdensity_EE}  \vspace{-0.3cm}
\end{figure}

\begin{figure}[t!]
\begin{center}
\includegraphics[width=\columnwidth]{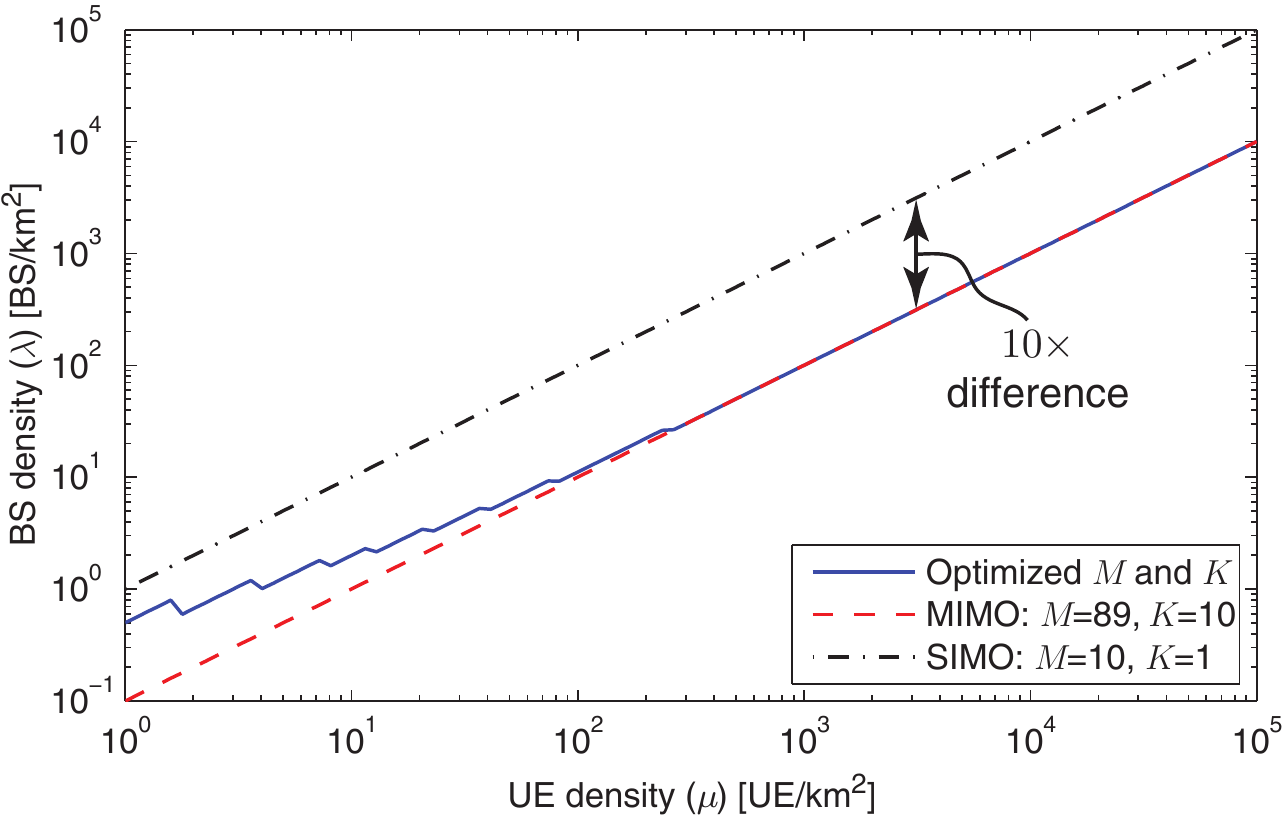}
\end{center}\vskip-4mm
\caption{Optimized BS density (in $\mathrm{BS/km}^2$) as a function of the UE density (in $\mathrm{UE/km}^2$). The system is optimized in the same way as in Fig.~\ref{figureUEdensity_EE}.} \label{figureUEdensity_BS} \vspace{-0.4cm}
\end{figure}

Fig.~\ref{figureUEdensity_EE} shows the EE as a function of the UE density $\mu $ for the average SINR level $\gamma = 3$, while Fig.~\ref{figureUEdensity_BS} shows the corresponding BS density. Apart from the optimal solution, we consider two reference cases: single-user single-input multiple-output (SIMO) transmission with $(M,K) = (10,1)$; and massive MIMO with $(M,K) = (89,10)$.

We notice that the EE level is roughly independent of the UE density for $\mu \geq 100$, which includes all the cases predicted in \cite{METIS_D11_short}. Single-user transmission is attractive for very low UE densities, while massive MIMO transmission is the best choice in the practical cases: Fig.~\ref{figureUEdensity_EE} shows how it improves the EE by a factor $3$ and Fig.~\ref{figureUEdensity_BS} shows that for a given UE density we can have $10 $ times fewer BSs per km$^2$. The latter has the benefit of greatly reducing the deployment costs.

\section{Conclusion}

\vspace{1mm}

Massive MIMO and small cells are two approaches to improve the energy efficiency of wireless networks. The main benefit of small cells is the reduced propagation losses, while the main benefit of massive MIMO is the interference suppression among the UEs that share the energy costs associated with the serving BS. Hence, these techniques are not competing but have complementary features that should be combined to achieve the maximal EE in future wireless networks; \emph{massive MIMO and small cells are a perfect match for marriage.}

\vspace{-0.2mm}

\bibliographystyle{IEEEbib}
\bibliography{IEEEabrv,refs}

\end{document}